# Quantum projection ghost imaging

**De-Zhong Cao (曹德忠)[1], Su-Heng Zhang (张素恒) [2*], Ya-Nan Zhao(赵亚楠) [2], Cheng Ren(任承)[1], Jun Zhang(张骏)[1], Baolai Liang(梁宝来)[2], Baoqing Sun(孙宝清)[3], and Kaige Wang (汪凯戈)[4**]**

[1]Department of Physics, Yantai University, Yantai 264005, Shandong Province, China

2College of Physics Science & Technology, Hebei University, Baoding 071002, Hebei Province, China

[3]School of Information Science and Engineering, Shandong University, Qingdao, 266237, Shandong Province, China

[4]Department of Physics, Applied Optics Beijing Area Major Laboratory, Beijing Normal University, Beijing 100875, China

*Corresponding author: hzhang@hbu.edu.cn; ** corresponding author: wangkg@bnu.edu.cn

We establish a quantum theory of computational ghost imaging and propose quantum projection imaging where object information can be reconstructed by quantum statistical correlation between a certain photon number of bucket signal and DMD random patterns. The reconstructed image can be negative or positive depending on the chosen photon number. In particular, the vacuum state (zero-number) projection produces a negative image with better visibility and contrast-to-noise ratio. The experimental results of quantum projection imaging agree well with theoretical simulations and show that, under the same measurement condition, vacuum projection imaging is superior to conventional and fast first-photon ghost imaging in low-light illumination.

**Keywords**: Ghost imaging, quantum projection, image reconstruction.

## I. Introduction

In recent years, low-photon flux imaging technology has attracted extensive attention due to its wide application in various fields, such as remote sensing, biological science and medical inspection, etc. [1-16]. The light source illuminating the object in most imaging systems is classical one, such as coherent light or thermal light. Under low-light illumination, only one or a few photons transmitted or reflected by the object can reach the detector with a certain probability. There are two major challenges. First, the zero-count of photons dominates the measurement process. In the intensity correlation only detected photons act as valid signal, and all vacuum signals are discarded. This greatly reduces the detection efficiency in imaging. Second, the shot noise in the classical source will lower signal-to-noise ratio (SNR) of the image. Quantum illumination in the imaging system, such as a single-photon source or a two-photon entangled source, would avoid these problems [16].

The development of the single-photon detector device and photon-counting techniques provides favorable conditions for low-light imaging. Kirmani et al. [2] proposed a low-flux imaging technique, called first-photon imaging, where for each pixel the number of illumination pulses prior to the first photon detection is used as an initial reflectivity estimate. They claimed that this avoids the Poisson noise inherent in low-flux operation. The similar way of first photon detection, called fast first-photon ghost imaging (FFPGI), can be applied to computational ghost imaging [8]. Sonnleitner et al. [3] utilized an image retrodiction approach that provides the full probability treatment for high-quality imaging data. Morris et al. [4] investigated how to obtain higher-quality images with a small number of photons. The authors attempted to find the answer to the question of how many photons does it take to form an image. Under using compressive techniques [14,15], they acquired the reconstructed image of a wasp wing with an average photon-per-pixel ratio of 0.45. A recent paper by Johnson et al. asked the same question as the title of their paper[10]. They pointed out a basic fact, that “for intensity images, it seems that one detected photon per image pixel is a realistic guide, but this may be reduced by making further assumptions on the sparsity of an image in a chosen basis, such as spatial frequency.” In any case, experimental results and experience tell us that under low-light conditions, increasing certain number of photons will significantly improve the image quality.

Under low-light illumination of a classical source, the quantum effects of light are revealed. In this paper, we set up the quantum theory of computational ghost imaging and propose quantum projection ghost imaging (QPGI). In the quantum projection scheme, the image is reconstructed by a certain number of photons of the object beam correlated with the random patterns of the digital micromirror device (DMD), and it thus can essentially avoid Poisson noise and improve image quality. Of particular interest is that quantum projection of the vacuum state can greatly increase imaging efficiency in the low-photon situation and still obtain a high-quality negative image as well.

## II. Theory

### A. General quantum theory for computational imaging

We first consider the quantum theory of computational ghost imaging. Computational ghost imaging usually relies on a digital micromirror device (DMD), which is an array of many micromirrors that flip independently. Each micromirror has two states, 0 and 1, which indicate that incident light is blocked or reflected. While a computer makes independently random control to all micromirrors, the probability function of each micromirror follows Bernoulli distribution

$$p(\alpha)=\begin{cases}1-q, & \alpha=0\\ q, & \alpha=1\end{cases}, \qquad (1)$$

where $\alpha$ is the binary variable and $q$ is the probability that micromirror is opened ($\alpha=1$).

Assume that each micromirror is illuminated by an independently and identically statistical source with the probability distribution $p_1(n)$ of photon number $n$. The joint-probability distribution of outgoing light for each micromirror is given by

$$p(\alpha,n)=\begin{cases}(1-q)\delta(n), & \alpha=0\\ qp_1(n), & \alpha=1\end{cases}, \qquad (2)$$

where $\delta(n)=1$ for $n=0$ and $\delta(n)=0$ for $n=1,2,\cdots$. It reflects a statistical correlation between the two random variables, $\alpha$ and $n$. If any number of photons join in the bucket detection, the micromirror must be opened. Otherwise the micromirror is closed with the probability of $1-q$. This is the key of computational ghost imaging. The photon probability distribution of outgoing light for each micromirror is obtained to be

$$p^{(1)}(n)=\sum_{\alpha} p(\alpha,n)=(1-q)\delta(n)+qp_1(n)\,. \qquad (3)$$

In comparison with $p_1(n)$ of incident light, the probability of the vacuum state is increased since the 0-state of the micromirror is taken into account. For $M$ independent micromirrors, the photon probability distribution can be derived by the convolutions

$$p^{(M)}(n)=p^{(1)}(n_1)\otimes p^{(1)}(n_2)\otimes\cdots\otimes p^{(1)}(n_M)$$

$$=(1-q)^M\left[\delta(n)+\sum_{l=1}^{M}C_M^l\left(\frac{q}{1-q}\right)^l p_l(n)\right], \qquad (4)$$

where $p_l(n)=p_1(n_1)\otimes\cdots\otimes p_1(n_l)$, $n=n_1+n_2+\cdots+n_M$, and $C_M^l=M!/(M-l)!l!$. Equation (4) is also valid for the case when a uniform source illuminates all DMD mirrors, where $p_l(n)$ is the photon probability distribution of any $l$ mirrors illuminated. For the common case of coherent light illumination, the two cases are equivalent. If $p_1(n)$ is a Poisson distribution with mean photon number $N_1$, $p_l(n)$ also satisfies the Poisson distribution with mean photon number $lN_1$. For simplicity, we consider a binary object, and all the photons transmitted or reflected from the object are received by the detector, the so-called bucket detection. Suppose the object is illuminated by $M$ micromirrors, i.e., the object has $M$ pixels. Equation (4) is the photon probability distribution for the bucket signals.

In a computational imaging scheme where the input light is coherent one with Poisson photon statistics (see details in Sec. Experiment), Figs. 1 (a1) and (a2) show the probability distribution of the bucket photons for the two cases. The probability of the vacuum state is clearly separated far from the other photon states because the 0-level of micromirrors only provides the probability of the vacuum state. The experimental results fit well with the theory curve, except for the 0 and 1 photon states, due to the stray light and dark counting of the single-photon avalanche detector (SPAD).

According to the mechanism of computational ghost imaging, for a given binary object all micromirrors are divided into two sets: on-micromirrors are imaged onto the object and off-micromirrors never do. The correlation of the bucket signal with the random variable of micromirrors will distinguish on-micromirror from off-one, and an image is formed. Hence we consider the joint probability function of two sets of micromirrors with the bucket signal. Any off-micromirror does not correlate with photons in bucket detection, so their joint probability function is written as

$$P_{\text{off}}(\alpha,n)=p^{(M)}(n)p(\alpha)\,. \qquad (5)$$

For an on-micromirror, the joint-probability distribution of outgoing light is described by Eq. (2). However, any one of on-micromirrors does not correlate with all others. As a result, the joint-probability distribution for the on-micromirror is given by

$$P_{\text{on}}(\alpha,n)=\sum_{n_1=0}^{n} p^{(M-1)}(n-n_1)p(\alpha,n_1)$$

$$=\begin{cases}(1-q)p^{(M-1)}(n), & \alpha=0\\ -(1-q)p^{(M-1)}(n)+p^{(M)}(n), & \alpha=1\end{cases}. \qquad (6)$$

The average photon number of outgoing beam from the pixel is $qN_1=\sum np^{(1)}(n)$, and the average photon number of the bucket signal for $M$ pixels is $\langle n\rangle_M=\sum np^{(M)}(n)=MqN_1$.

The above Eqs. (5) and (6) summarize the basic quantum description of computational ghost imaging. So now we can calculate the correlation coefficients for imaging. We first define the mean value and mean square value of photon number injected onto each micromirror as

$$N_1\equiv\sum np_1(n)\,, \qquad (7a)$$

$$N_2\equiv\sum n^2p_1(n)\,. \qquad (7b)$$

Using Eq. (3), the corresponding mean values of outgoing photons for the micromirror are

$$\langle n\rangle\equiv\sum np^{(1)}(n)=qN_1\,, \qquad (8a)$$

$$\langle n^2\rangle\equiv\sum n^2p^{(1)}(n)=qN_2\,. \qquad (8b)$$

Since all micromirrors are independent of each other, the mean value and mean square value of bucket photons are obtained to be

$$\langle n\rangle_M\equiv\sum np^{(M)}(n)=MqN_1\,, \qquad (9a)$$

$$\langle n^2\rangle_M\equiv\sum n^2p^{(M)}(n)=\langle(n_1+n_2+\cdots+n_M)^2\rangle$$

$$=Mq[N_2+(M-1)qN_1^2]. \qquad (9b)$$

The first-order correlation coefficients between the binary variable and bucket photon number for off- and on-micromirrors are calculated to be

$$\langle\alpha n\rangle_{\text{off}}=\sum_{\alpha,n}\alpha nP_{\text{off}}(\alpha,n)=q^2MN_1\,, \qquad (10)$$

$$\langle\alpha n\rangle_{\text{on}}=\sum_{\alpha,n}\alpha nP_{\text{on}}(\alpha,n)=\sum_{n}[np^{(M)}(n)-n(1-q)p^{(M-1)}(n)]$$

$$=qMN_1-(1-q)q(M-1)N_1=q^2MN_1+q(1-q)N_1\,. \qquad (11)$$

Hence imaging pixels can be distinguished from non-imaging pixels. The imaging visibility is obtained as

$$V=\frac{\langle\alpha n\rangle_{\text{on}}-\langle\alpha n\rangle_{\text{off}}}{\langle\alpha n\rangle_{\text{on}}+\langle\alpha n\rangle_{\text{off}}}=\frac{1}{1+2Mq/(1-q)}\,. \qquad (12)$$

The expression is general, independent of the intensity and statistical nature of the incident light. When $q=1/2$, $V=1/(1+2M)$ is the same as the visibility of ghost imaging with thermal light correlation. Therefore computational ghost imaging can greatly improve visibility by decreasing the probability of $q$. The price paid is a reduction in detection efficiency.

The second-order correlation coefficients for off- and on-micromirrors are written as

$$\langle\alpha^2n^2\rangle_{\text{off}}=\sum\alpha^2p(\alpha)\cdot\sum n^2p^{(M)}(n)=q\cdot\langle n^2\rangle_M\,, \qquad (13)$$

$$\langle\alpha^2n^2\rangle_{\text{on}}=\sum n^2p^{(M)}(n)-(1-q)\sum n^2p^{(M-1)}(n)$$

$$=\langle n^2\rangle_M-(1-q)\langle n^2\rangle_{M-1}\,. \qquad (14)$$

With these coefficients we can calculate the contrast-to-noise ratio (CNR)[18]

$$\text{CNR}=\frac{|\langle\alpha n\rangle_{\text{on}}-\langle\alpha n\rangle_{\text{off}}|}{\sqrt{(\langle\alpha^2n^2\rangle_{\text{off}}-\langle\alpha n\rangle_{\text{off}}^2)+(\langle\alpha^2n^2\rangle_{\text{on}}-\langle\alpha n\rangle_{\text{on}}^2)}}\,. \qquad (15)$$

## B. Quantum projection imaging

Here we propose a novel imaging scheme using quantum projection detection. As has shown in Figs. 1 (a1) (a2), the bucket photons contain great fluctuations. To bypass the fluctuations, we measure and pick a specific number of photons in the bucket signal and corresponding DMD pixel patterns to form the image.

From the probability functions (5) and (6) of computational imaging we calculate the corresponding conditional probability distributions for a given photon number $k$ :

$$P_{\text{off}}^{(c)}(\alpha\,|\,k) = p(\alpha), \qquad (16)$$

$$P_{\text{on}}^{(c)}(\alpha\,|\,k) = P_{\text{on}}(\alpha,k)\big/ p^{(M)}(k)$$

$$= \begin{cases} (1-q)\,p^{(M-1)}(k)\,/\,p^{(M)}(k), & (\alpha = 0) \\ 1-(1-q)\,p^{(M-1)}(k)\,/\,p^{(M)}(k). & (\alpha = 1) \end{cases} \qquad (17)$$

With the fact $\alpha = \alpha^2$ , the first- and second-order conditional correlation coefficients are obtained equally to be

$$\langle \alpha | k \rangle_{\text{off}} = \langle \alpha^2 | k \rangle_{\text{off}} = \sum \alpha\, p(\alpha) = q, \qquad (18)$$

$$\langle \alpha | k \rangle_{\text{on}} = \langle \alpha^2 | k \rangle_{\text{on}} = \sum\nolimits_{\alpha} \alpha P_{\text{on}}^{(c)}(\alpha\,|\,k)$$

$$= 1-(1-q)\,p^{(M-1)}(k)\,/\,p^{(M)}(k).. \qquad (19)$$

The visibility of quantum projection imaging depends on the measurement of $k$ photons:

$$V(k) = \frac{\langle \alpha | k \rangle_{\text{on}} - \langle \alpha | k \rangle_{\text{off}}}{\langle \alpha | k \rangle_{\text{on}} + \langle \alpha | k \rangle_{\text{off}}}$$

$$= \frac{(1-q)[p^{(M)}(k) - p^{(M-1)}(k)]}{[p^{(M)}(k) - p^{(M-1)}(k)] + q[p^{(M)}(k) + p^{(M-1)}(k)]} \qquad .(20)$$

The significant feature of quantum projection imaging compared to conventional is its potential to produce negative images. The condition for the formation of a negative image is

$$p^{(M)}(k) < p^{(M-1)}(k). \qquad (21)$$

In this case the object signal $\langle \alpha | k \rangle_{\text{on}}$ is less than the background $\langle \alpha | k \rangle_{\text{off}}$ . The similar effects of positive and negative ghost imaging by conditional collection of certain range of bucket intensities have been reported both experimentally and theoretically [19-22].

Visibility of quantum projection imaging can be expressed as the similar form as Eq. (12):

$$V(k) = \frac{1}{1 + 2\mu(M,k)q\,/\,(1-q)}, \qquad (22)$$

where the effective value of object pixels is defined as

$$\mu(M,k) \equiv \frac{1}{1 - p^{(M-1)}(k)\,/\,p^{(M)}(k)}. \qquad (23)$$

The positive and negative values of $\mu(M,k)$ correspond to the positive and negative images, respectively. For $0 < \mu(M,k) < M$ , quantum projection positive imaging has better visibility than the conventional one. As for the negative values of $\mu(M,k)$ , however, under the condition of

$$(1-q)\,/\,q < -\mu(M,k) < M + (1-q)\,/\,q, \qquad (24)$$

negative imaging has better visibility than conventional one. The visibility reaches the perfect value when $-\mu(M,k) \to (1-q)\,/\,q$ .

CNR of quantum projection imaging is defined as

$$\text{CNR} = \frac{\left| \langle \alpha | k \rangle_{\text{on}} - \langle \alpha | k \rangle_{\text{off}} \right|}{\sqrt{\left( \langle \alpha^2 | k \rangle_{\text{off}} - \langle \alpha | k \rangle_{\text{off}}^2 \right) + \left( \langle \alpha^2 | k \rangle_{\text{on}} - \langle \alpha | k \rangle_{\text{on}}^2 \right)}}. \qquad (25)$$

Applying Eqs. (18) and (19) to it, we obtain

$$\text{CNR} = \frac{\sqrt{1-q}\,|\,1 - p^{(M-1)}(k)\,/\,p^{(M)}(k)\,|}{\sqrt{q + [1-(1-q)\,p^{(M-1)}(k)\,/\,p^{(M)}(k)]\,p^{(M-1)}(k)\,/\,p^{(M)}(k)}}$$

$$= \sqrt{\frac{1-q}{q(2\mu^2 - 2\mu + 1) + \mu - 1}}, \qquad (26)$$

where $\mu$ has been defined by Eq. (23).

In the same experiment associated with Figs. 1(a1) (a2), Figs. 1(b1) (b2) show the effective value of the object pixels $\mu(M,k)$ versus the projection photon number $k$ of the bucket signal, and Figures 1(c1) (c2) and (d1) (d2) show corresponding visibility and CNR of quantum projection imaging, respectively. In Fig. 1(b1), we can see that the positive imaging and negative imaging exist when the projection photon number $k \geq 14$ and $k \leq 13$ , respectively. In the case of positive imaging, $\mu < M$ is valid for a larger $k$ ($k > 23$), and it gets the better visibility and CNR [see (c1) and (d1)]. However, the vacuum and small photon number ($k < 10$) projection in the negative imaging case also have the better visibility and CNR than the conventional imaging, as long as Eq. (24) is satisfied. Especially for the vacuum projection, $-\mu(M, k=0) \to (1-q)\,/\,q$ , the visibility is almost perfect and CNR is much higher than the others.

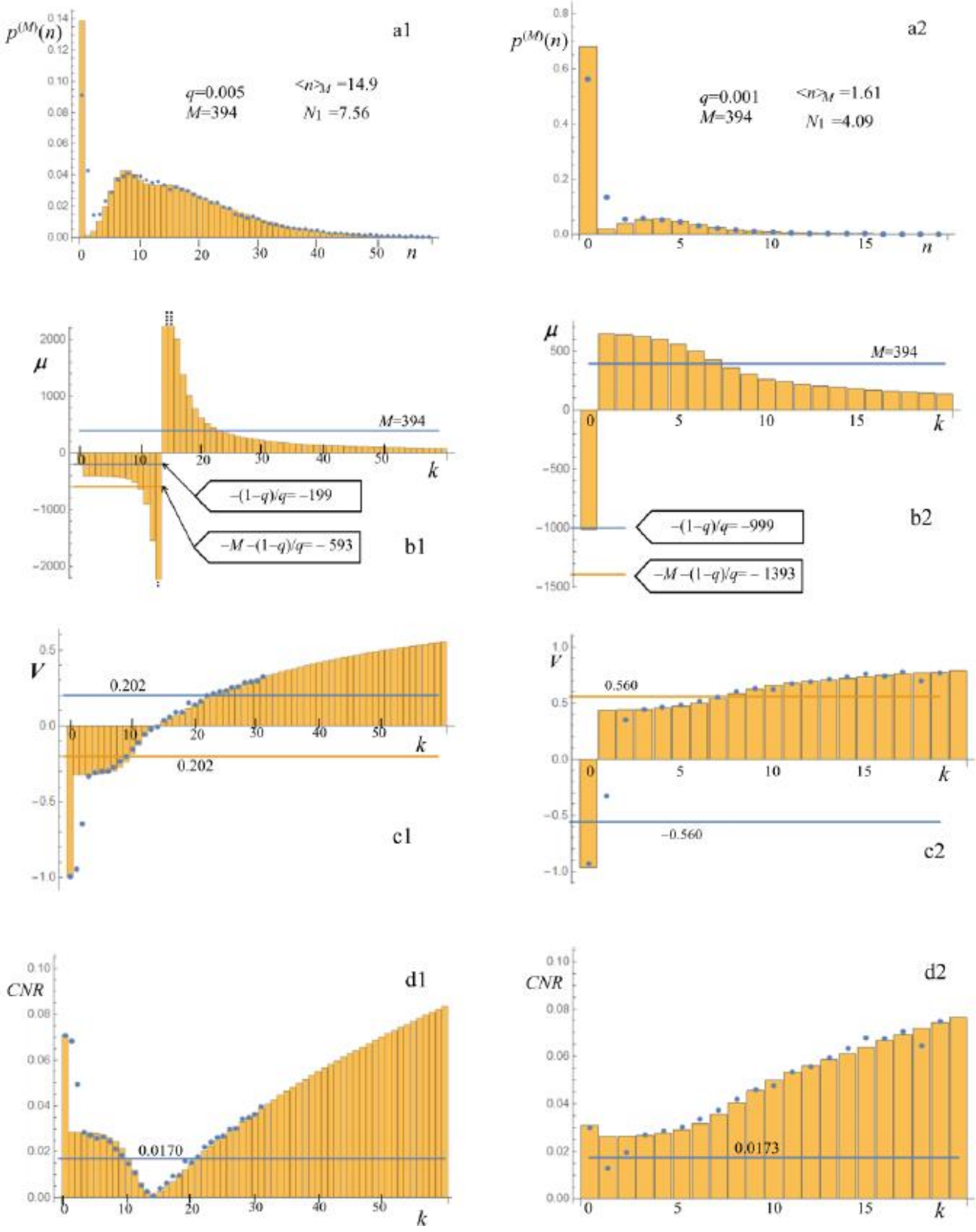

**Fig. 1** Photon probability distributions of the bucket signals of an object ( $M = 394$ ) for the two cases: (a1) $q = 0.005$ , $N_1 = 7.56$ and (a2) $q = 0.001$ , $N_1 = 4.09$ . Corresponding effective values of object pixels, visibilities and contrast-to-noise ratios (CNR) of quantum projection imaging are shown in (b1)(b2), (c1)(c2), (d1)(d2), respectively. Barcharts are the theoretical simulations and circles are the experimental results. In (b1)(b2), the two lines under the horizontal axis show the range defined by Eq. (24). In (c1)(c2) and (d1)(d2), the lines indicate the corresponding values of conventional ghost imaging for comparison.

According to Eq. (4), it can be proved that when $Mq < 1$ the pixel number of the illuminated object is less than 1, all the quantum projection images are positive except the vacuum projection imaging. Figures 1(a2)-(d2) show the experiment for the case $M=394$ and $q=0.001$, which satisfies $Mq<1$. The vacuum projection imaging has much better visibility than other ways. In most cases, however, CNR of quantum projection imaging is better than that of conventional imaging.

It is worth pointing out that in computational ghost imaging when some frames are selected for imaging, the remaining frames can be superimposed to form a complementary image[22]. In quantum imaging of the $k$-photon projection, the corresponding complementary image of non-$k$-photon projection is formed by projecting all other photon numbers except the $k$-photon. The probability of bucket detection of non $k$-photon projection is given by $p^{(M)}(\bar{k})=1-p^{(M)}(k)$. Replacing $p^{(M)}(k)$ with $p^{(M)}(\bar{k})$ in Eqs. (17), (22), (23) and (26), we obtain the visibility and CNR for the complementary image. According to the mathematical formulae, there is in general no simple relationship between a pair of complementary images in terms of the visibility and CNR.

### C. Vacuum projection imaging

In the quantum projection scheme above, a very peculiar option is the vacuum projection, where no photons are detected in the bucket signal. Let $p_1(0)$ be the probability of the vacuum state for the incident light on each micromirror. The corresponding vacuum probability for the outgoing light of the micromirror is $p^{(1)}(0)=1-q+qp_1(0)$. Since all micromirrors are statistically independent of each other, for $M$ micromirrors it has $p^{(M)}(0)=[p^{(1)}(0)]^M=[1-q+q\cdot p_1(0)]^M$. Therefore, the condition of negative image $p^{(M)}(0)<p^{(M-1)}(0)$ is always satisfied for vacuum projection detection. According to Eq. (20), the visibility of vacuum projection imaging is obtained to be

$$V(k=0)=-\frac{(1-q)[1-p_1(0)]}{1-q+(1+q)p_1(0)}. \qquad (27)$$

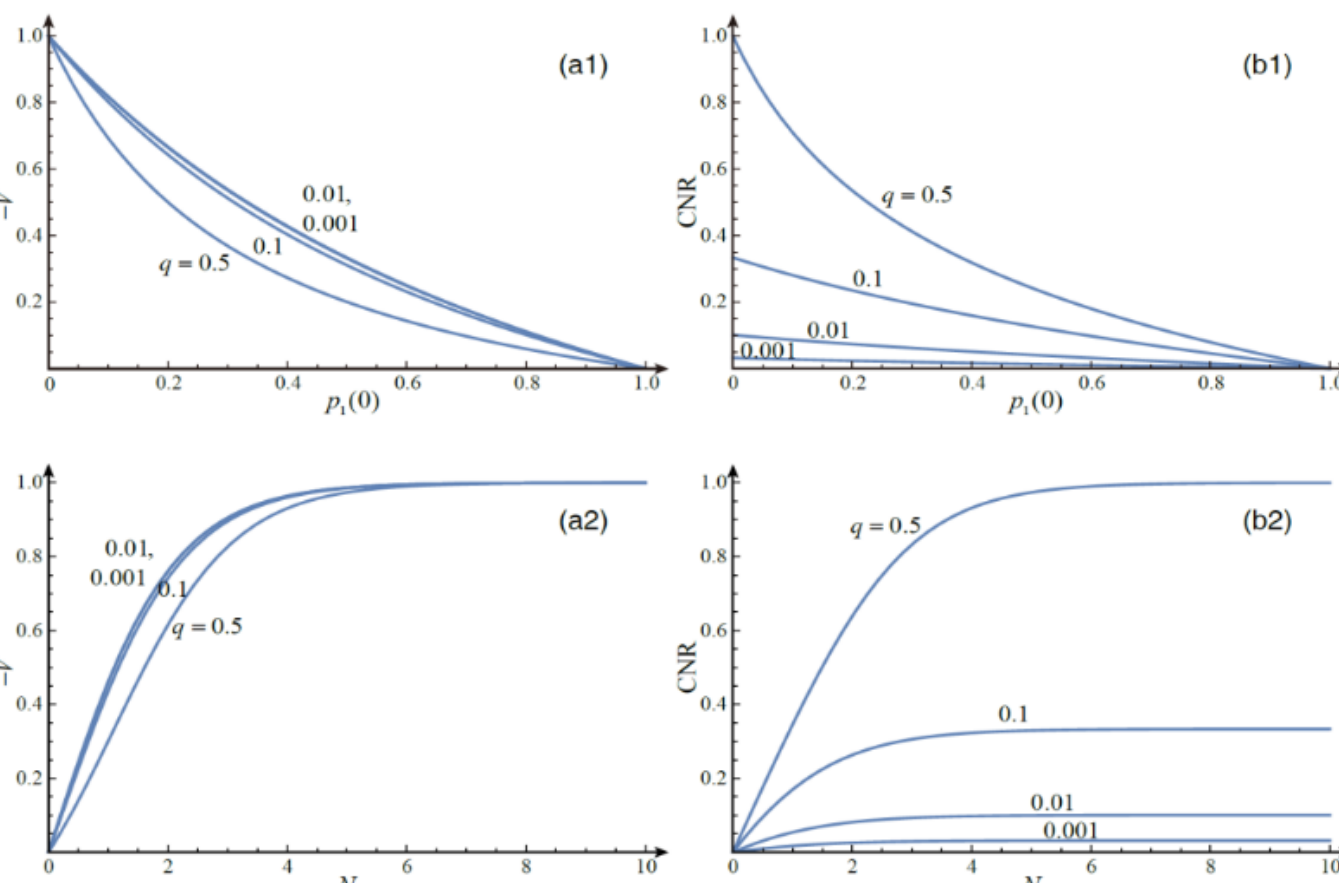

**Fig. 2** Visibilities (a1)(a2) and CNRs (b1)(b2) for vacuum projection imaging, where $p_1(0)$ is the vacuum probability for outgoing light of a micromirror, and $N_1$ is the corresponding average photon number for the Poisson distribution.

The corresponding CNR for vacuum projection imaging is written as

$$\mathrm{CNR}=\frac{|\langle\alpha|0\rangle_{\mathrm{on}}-\langle\alpha|0\rangle_{\mathrm{off}}|}{\sqrt{(\langle\alpha^2|0\rangle_{\mathrm{off}}-\langle\alpha|0\rangle_{\mathrm{off}}^2)+(\langle\alpha^2|0\rangle_{\mathrm{on}}-\langle\alpha|0\rangle_{\mathrm{on}}^2)}}$$

$$=[1-p_1(0)]\sqrt{\frac{q(1-q)}{[1-q+qp_1(0)]^2+p_1(0)}}. \qquad (28)$$

For the coherent light with average photon number $N_1$, $p_1(0)=\exp[-N_1]$. In the vacuum projection cases, imaging visibility and CNR versus the probability of vacuum state are shown in Figs. 2 (a1) and (b1), and versus the average photon number $N_1$ in Figs. 2 (a2) and (b2), respectively. Figure 2 tells us that appropriate intensity of driving light ensures good imaging result of vacuum projection. In the case of a larger average photon number $N_1$, a small probability $q$ can be selected to maintaina large probability of the vacuum state as shown in Fig. 1.

As indicated in Eqs. (12) and (15) above, the visibility and CNR of computational ghost imaging are inversely proportional to the object pixels. In quantum projection imaging, however, the simple inverse relationship no longer holds. Interestingly, for vacuum projection imaging, Eqs. (27) and (28) show that both visibility and CNR are completely independent of object pixels. This feature will greatly improve the image quality and resolution in the face of large and complex objects in computational ghost imaging.

## III. Experiment

The experimental setup is similar to conventional computational ghost imaging but with a photon counting system replacing the intensity detection, as shown in Fig. 3. The photon counting system consists of a SPAD (Excelitas SPCM-AQRH-W6) and a time-correlated single-photon counting TCSPC module (PicoQuant PicoHarp 300) with a time resolution of 4 picoseconds. The optical source is a super continuum pulsed laser SCPL (NKT: SuperK EXTREME) with a temporal pulse width of 20 picoseconds and a frequency of 6.49 MHz. After passing through a filter, a beam of wavelength 660 nm is selected to illuminate the object. The laser beam is strongly attenuated by a neutral density filter (NDF: Daheng GCC-3010) before hitting DMD (Xintong F4100). The object beam is reflected by DMD and registered by a single-pixel SPAD with the help of a collecting lens L2. The DMD contains 1,024×768 independently addressable micromirrors, and is used to load random patterns and/or virtual objects (a music note and resolution bars).

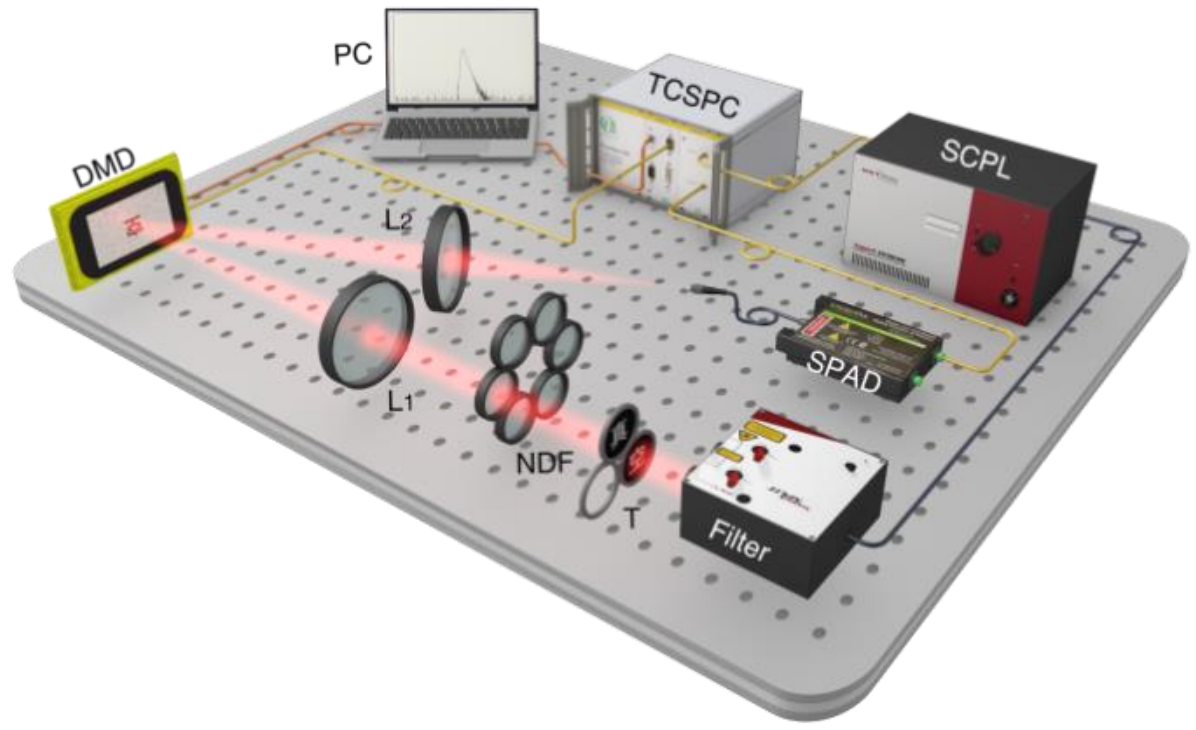

**Fig. 3** Experimental setup of quantum projection imaging. SCPL represents the super continuum pulsed laser. Filteris used to select out the beam wavelength of 660nm. T is an object and NDF is the neutral density filter. L1 and L2 are the imaging lens and collective lens, respectively. DMD is the digital micromirror device. SPAD is the single-photon avalanche detector. TCSPC is the module of time-correlated single-photon counting. PC is a computer.

The refresh time of the DMD frame can be set in the range of $5\times10^{-4}$ s~ $2\times10^{-3}$ s. For example, if the refresh time is $10^{-3}$ s, the corresponding repetition rate of each frame is 6,490 for the pulse frequency 6.49M of the driving laser. We record the total number of photons in these 6,490 pulses as the photon counts of the frame. For best experimental results, the photon number measured in the frame must be much smaller than the repetition rate of each frame.

We first create a virtual object in DMD, a musical note with $M=394$ pixels, and consider the two cases of the probabilities that the micromirror is opened, $q=0.005$ and $q=0.001$. We measure the photon statistical distributions of the bucket signal, where the average photon numbers are $\langle n\rangle_M=14.9$ and $\langle n\rangle_M=1.61$ as shown in Figs. 1 (a1) and (a2), respectively. All the experimental results are shown with the circles and consistent well with the theoretical simulation of Eq. (4) for the Poisson photon distribution of the driving beam. In the experimental results, as have already pointed out above, the decrease in 0-photon count and the increase in 1- and 2-photon counts are due to the stray light and dark counting of SPAD.

Figures 1(c1) (c2) and Figs. 1(d1) (d2) are the visibility and CNR for the two cases of quantum projection imaging, respectively. In Fig. 1(c1), the theoretical and experimental results show that negative images occur when the number of projection photons $k$ is not greater than 13, otherwise positive images occur. The theoretical derivation manifests that when $Mq<1$ negative imaging can only exist in vacuum projection, and this is verified in Fig. 1(c2). However, the experimental result shows the negative visibility for $k$=1-photon. This inconsistency is due to the fact that as many 0-photon counts have been added into the 1-photon case, as has been shown in Fig. 1(a2). The experimental results of CNR are shown in Figs. 1(d1) (d2). We can see that CNR and visibility change synchronously. In general, all the experimental results in Figs. 1 are in good aggrement with the theoretical curves.

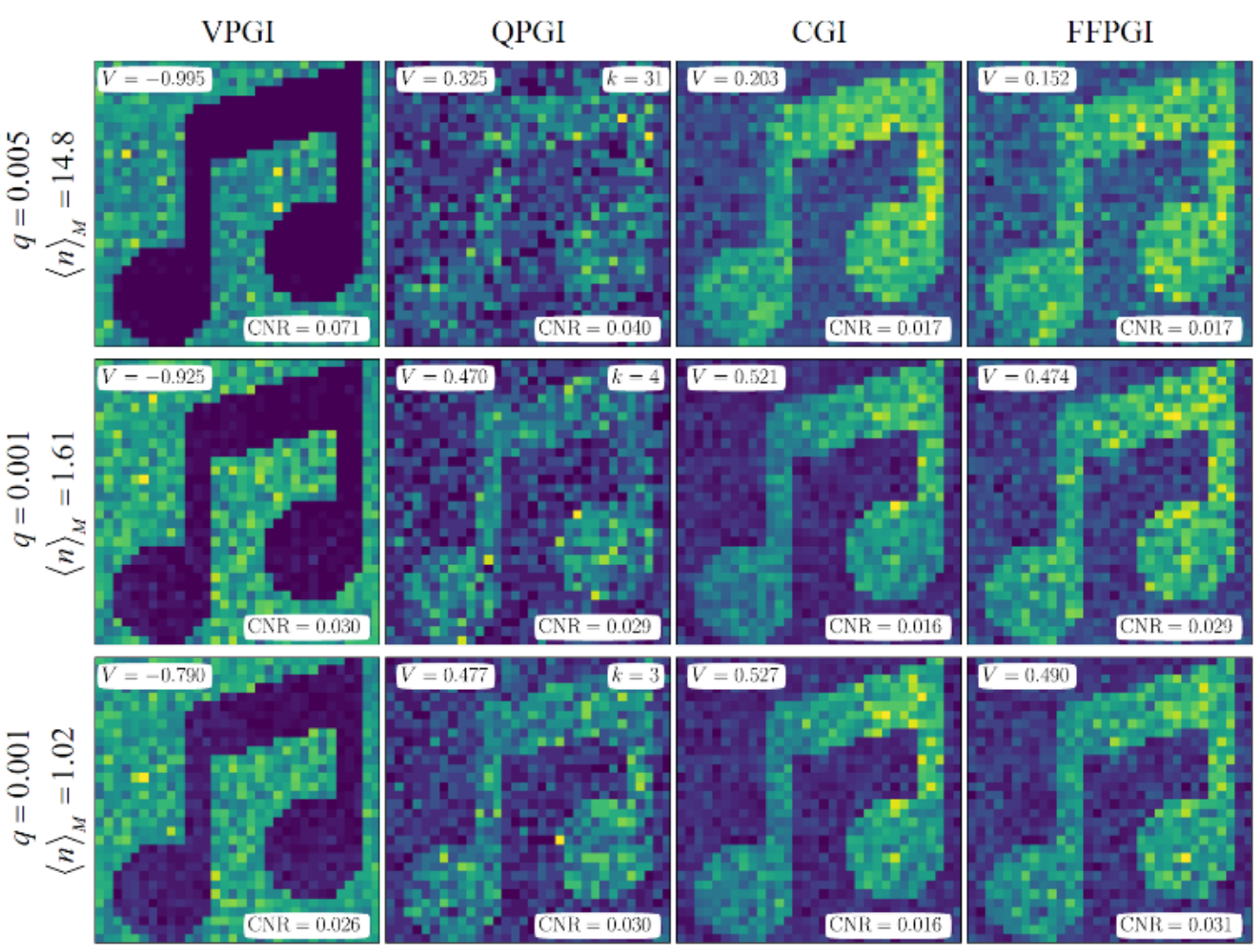

**Fig. 4** Reconstructed images of a musical note are observed with four ways, VPGI, QPGI with $k$-photon, CGI and FFPGI, in the three cases of average photon numbers $\langle n\rangle_M$ and probabilities $q$.

In conjunction with Fig. 1, Fig. 4 shows the experimental observation of reconstructed images for a virtual object (music note) of $M=394$ pixels, and all values of imaging visibility and CNR are listed in the plots. For the three cases of average photon numbers $\langle n\rangle_M$ and probabilities $q$ of micromirror being opened, we measured the same number of 40,960 DMD frames and reconstructed the images in the four ways: vacuum projection ghost imaging (VPGI), quantum projection ghost imaging with $k$ photons (QPGI), conventional ghost imaging (CGI), and fast first-photon ghost imaging (FFPGI). In case (i) $q=0.005$ and $\langle n\rangle_M=14.8$, the frame numbers recorded in vacuum projection and $k=31$ photons projection are 3,729 and 432, respectively. In case (ii) $q=0.001$ and $\langle n\rangle_M=1.61$, the frame numbers recorded in vacuum projection and $k$=4 photons projection are 22,968 and 2,117, respectively. In case (iii) $q$=0.001 and $\langle n\rangle_M=1.02$, the frame numbers recorded in vacuum projection and $k=3$ photons projection are 25,535 and 2,549, respectively. Obviously, the negative images of vacuum projection are much better than all other positive images in terms of visibility and CNR.

In Fig. 5 the virtual objects are resolution bars in various combinations. We use the same probability $q=0.005$, the photon number $N_1=7.86$ of each micromirror, and 100,000 DMD frames for all the cases. The experimental results of vacuum projection ghost imaging (VPGI), conventional ghost imaging (CGI), and non-vacuum projection ghost imaging (NVPGI) for the same objects are plotted in the first, second and third columns, respectively. The corresponding photon number distributions of the bucket detection are plotted in the fourth column. It is clear that as the number of object pixles increases, both the visibility and CNR of VPGI can maintain high values unchanged, while both decrease in conventional ghost imaging and NVPGI (the complementary imaging of VPGI).

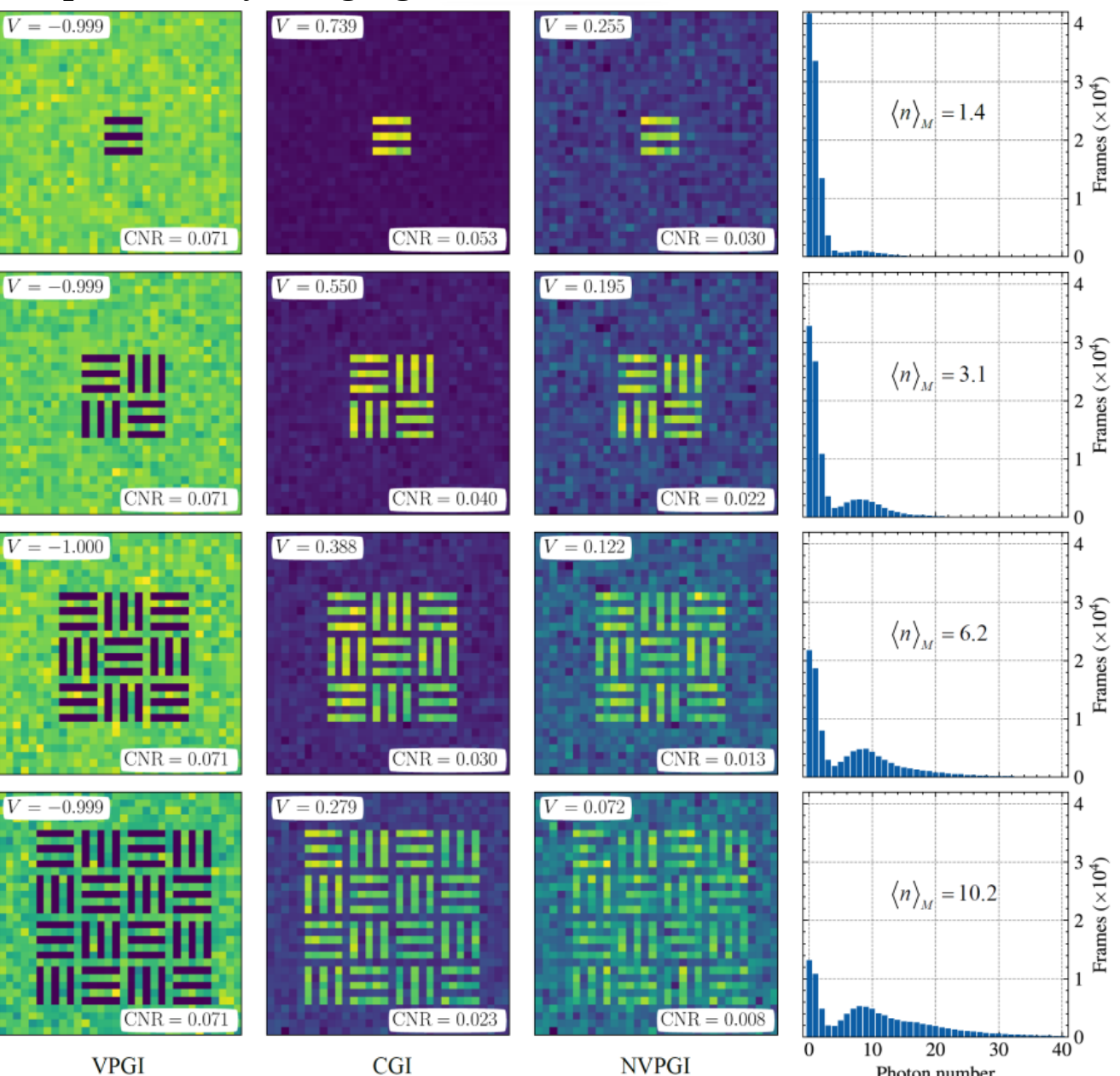

**Fig. 5** Reconstructed images of resolution bars in various combinations with VPGI (first column), CGI (second column) and NVPGI (third column). The corresponding photon number distributions of bucket detection are plotted in the fourth column.

Two Chinese characters "zhen" and "kong" (vacuum) as real objects are also used in the experimental scheme of Fig. 3. The experimental results and relevant data are shown in Figure 6. In all 40,000 DMD frames, 4,696 and 7,508 frames are assigned to the vacuum projection for "zhen" and "kong", respectively, and two high-quality negative images are formed with these frames. The reconstructed images of the real objects also demonstrate that VPGI is the best choice with respect to CGI and FFPGI.

## IV. Discussion and conclusion

Under strong enough light illumination, both conventional imaging and computational ghost imaging can easily produce the best results. However, in low-light conditions, increasing the exposure time or the number of exposures can significantly improve the image quality. That is, a sufficient number of photons is required to achieve a high-quality image. The answer to the question "how many photons does it take to form an image?" also depends on the quality of the image. Perhaps a more appreciate question to ask is: what way can observe the best image for a given total number of photons?

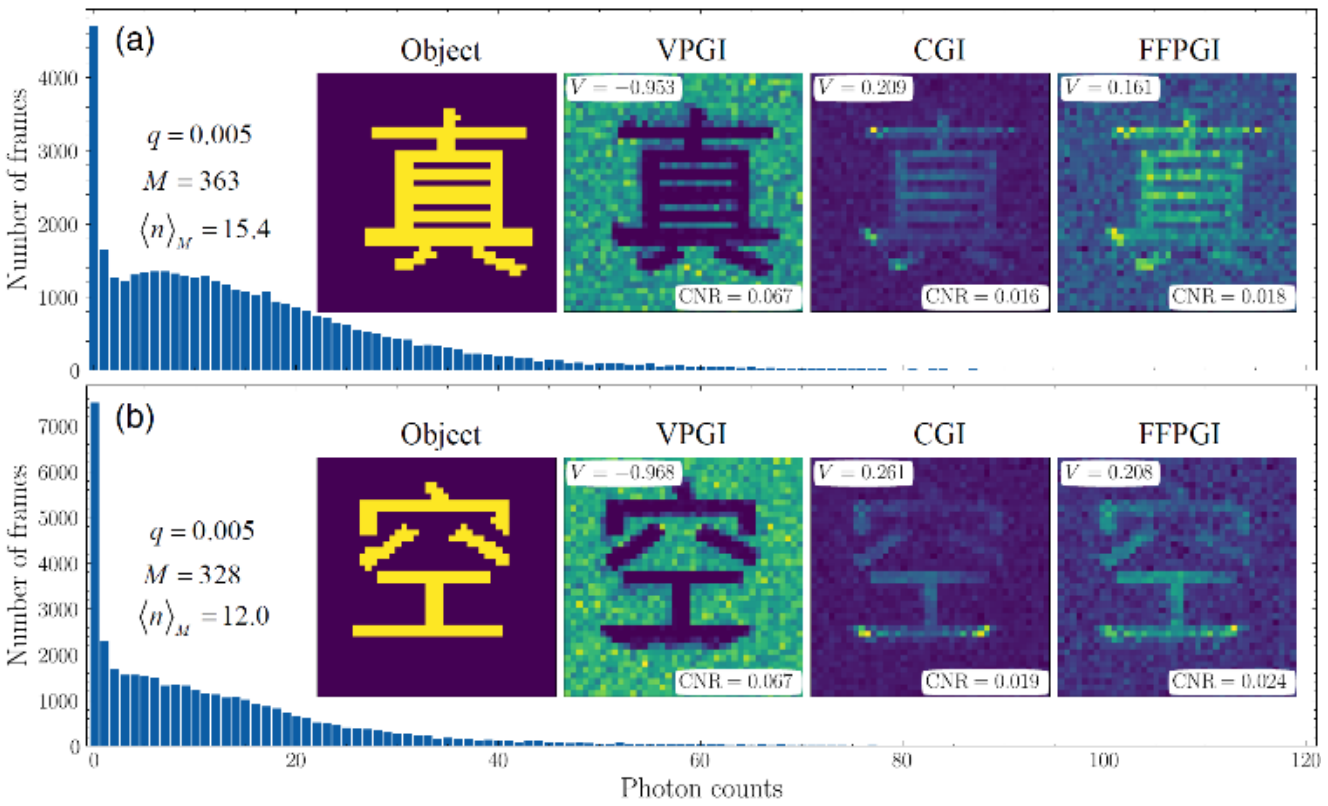

**Fig. 6** Number of frames versus photon counts and reconstructed images of two Chinese characters with VPGI, CGI and FFPGI in (a) "zhen" and (b) "kong". The total number of frames is 40,000.

For low-light imaging application, we establish the quantum theory of computational ghost imaging. Based on the quantum statistical correlation between the photon counts of the bucket signal and DMD random patterns, we propose quantum projection imaging in which reconstructed image is formed by detecting a particular photon count in the bucket signal. The vacuum state and lower-photon counting projections yield negative imaging while projections with higher photon counts yield positive imaging. Both theoretical and experimental results have shown that proper selection of quantum projection imaging can lead to better imaging results than conventional ghost imaging and fast first photon ghost imaging. Quantum projection imaging, as a pure quantum version applied to computational ghost imaging, will attract much attention. In particular, vacuum projection imaging can achieve the best negative image, especially its visibility and CNR are independent of object pixels. This important feature will greatly promote the wide application of quantum imaging technology in large and complex scenes.

## Acknowledgement

This work is financially supported by the National Natural Science Foundation of China (Grant No. 12274037, 11674273) and the Natural Science Foundation of Hebei Province (A202220103).